\documentclass{article}
\usepackage{subcaption}
\usepackage{spconf,amsmath,graphicx, amssymb,amsfonts}
\usepackage{algorithm}
\usepackage{algorithmic}
\usepackage{color,soul}
\usepackage{multirow}
\usepackage{graphicx}
\usepackage{booktabs}
\usepackage{textcomp}
\usepackage[dvipsnames,table]{xcolor}

\DeclareMathOperator*{\argmin}{arg\,min}

\newcommand{\algcolor}[2]{\hspace*{-\fboxsep}\colorbox{#1}{\parbox{\linewidth}{#2}}}
\newcommand{\algehlgy}[1]{\algcolor{blue!20}{#1}}
\newcommand{\algehlbg}[1]{\algcolor{red!30}{#1}}

\title{Federated cINN Clustering for Accurate Clustered Federated Learning}
\name{Yuhao Zhou$^1$, Minjia Shi$^1$, Yuxin Tian$^1$, Yuanxi Li$^2$, Qing Ye$^1$, Jiancheng Lv$^1$}
\address{$^1$~Sichuan University \\ $^2$~University of Illinois at Urbana-Champaign}
\begin{document}
%
\maketitle
\begin{abstract}
Federated Learning (FL) presents an innovative approach to privacy-preserving distributed machine learning and enables efficient crowd intelligence on a large scale. 
However, a significant challenge arises when coordinating FL with crowd intelligence which diverse client groups possess disparate objectives due to data heterogeneity or distinct tasks. 
To address this challenge, we propose the Federated cINN Clustering Algorithm (FCCA) to robustly cluster clients into different groups, avoiding mutual interference between clients with data heterogeneity, and thereby enhancing the performance of the global model.
Specifically, FCCA utilizes a global encoder to transform each client's private data into multivariate Gaussian distributions. 
It then employs a generative model to learn encoded latent features through maximum likelihood estimation, which eases optimization and avoids mode collapse. 
Finally, the central server collects converged local models to approximate similarities between clients and thus partition them into distinct clusters. 
Extensive experimental results demonstrate FCCA's superiority over other state-of-the-art clustered federated learning algorithms, evaluated on various models and datasets. These results suggest that our approach has substantial potential to enhance the efficiency and accuracy of real-world federated learning tasks.
\end{abstract}
\section{Introduction}

Federated Learning (FL) has made a promising entrance as an effective approach for various applications~\cite{kairouz2021advances} while complying with data privacy regulations such as GDPR\footnote{https://gdpr-info.eu/}, HIPAA\footnote{https://www.hhs.gov/hipaa/for-professionals/privacy/laws-regulations/index.html}, and CCPA\footnote{https://oag.ca.gov/privacy/ccpa}. However, the decentralized nature of FL can result in significant data heterogeneity~\cite{li2020federated}, which leads to divergent learning trajectories and inconsistencies in model performance~\cite{ghosh2020efficient}. To surmount this challenge, researchers have introduced clustered Federated Learning (clustered FL)~\cite{briggs2020federated,sattler2020clustered,ghosh2020efficient,long2023multi} to ensure that clients with similar data distributions collaborate to train the same model, thus alleviating the impact of data heterogeneity on the overall performance of the FL framework with crowd intelligence~\cite{yang2022crowd}. 

Nevertheless, existing clustered FL approaches~\cite{briggs2020federated,sattler2020clustered,ghosh2020efficient,long2023multi} group clients in each step based on the clustering solution of the prior steps. This can result in a cascade of errors and sub-optimal clustering solutions as the errors propagate through subsequent steps, particularly in the nascent phase of training. 

In this paper, we present a novel solution to the challenge at hand. We introduce the Federated cINN Clustering Algorithm (FCCA), which performs accurate clustered FL by means of a sophisticated architecture, consisting of a global encoder for representing clients' private data distributions, a conditional Invertible Neural Network (cINN)~\cite{ardizzone2019guided} for continuous learning of encoded representations without mode collapse, and a similarity assessment and clustering algorithm under extreme non-i.i.d. data, all without relying on previous clustering solutions. Extensive experiments conducted on various datasets demonstrate the benefits of FCCA over other clustered FL algorithms. The source codes of all experiments are open-sourced for reproduction.

\section{Related Work}
\subsection{Clustered Federated Learning}

Clustered FL~\cite{briggs2020federated,sattler2020clustered,ghosh2020efficient,long2023multi} groups clients based on their data distributions and trains models on these groups separately to mitigate the impact of data heterogeneity in FL. Although existing algorithms are effective in certain cases, they rely heavily on previous clustering solutions. To address this limitation, recent studies~\cite{kim2021dynamic} propose the use of Generative Adversarial Networks (GAN)~\cite{goodfellow2020generative} to consistently represent clients' local data for clustering while minimizing dependence on previous clustering solutions. However, direct learning from raw local data presents serious security and privacy concerns. In addition, GAN-based clustered FL requires substantial computational resources and is often plagued by practical issues such as mode collapse and convergence problems. In this paper, our main focus is to reduce the dependence on clustered FL while protecting user privacy and achieving high clustering accuracy.

\subsection{Conditional Invertible Neural Network}
Conditional Invertible Neural Networks (cINNs) are an exemplary class of INN-based~\cite{dinh2016density} class-conditional generative models that can generate complex data samples without the mode collapse issue and require fewer runtime resources. One keystone of INNs is the Affine Coupling Block, where the input $u$ is divided into $[u_1, u_2]$, and subsequently converted into $[v_1, v_2]$ by an affine transformation denoted as Equation~\ref{eq:acb-forawrd}.
\begin{equation}
\left\{
\begin{aligned}
v_1 &= u_1 \odot exp(s_1(u_2)) + t_1(u_2) \\
v_2 &= u_2 \odot exp(s_2(v_1)) + t_2(v_1)
\end{aligned}
\right.
\label{eq:acb-forawrd}
\end{equation}
Here, $s_k$ and $t_k$ can be any neural networks. Then, by inverting Equation~\ref{eq:acb-forawrd}, $[u_1, u_2]$ can be retrieved from $[v_1, v_2]$. Next, by conditioning the affine coupling block, cINNs establish invertibility through $f^{-1}(\mathbf{x}; \mathbf{c}, \theta^c) = g(\mathbf{z}; \mathbf{c}, \theta^c)$, where $f(\mathbf{x}; \mathbf{c}, \theta^c)$ is referred as a cINN network parameterized by $\theta^c$ and conditioning data $\mathbf{c}$ with input $\mathbf{x}$, and $g(\mathbf{z}; \mathbf{c}, \theta^c)$ is the inverse function of $f(\cdot)$. cINNs have shown promise in applications such as image completion, anomaly detection, and privacy-preserving solutions. 

\section{Federated cINN Clustering Algorithm}
\begin{figure}[htbp]
\centerline{\includegraphics[width=0.5\textwidth]{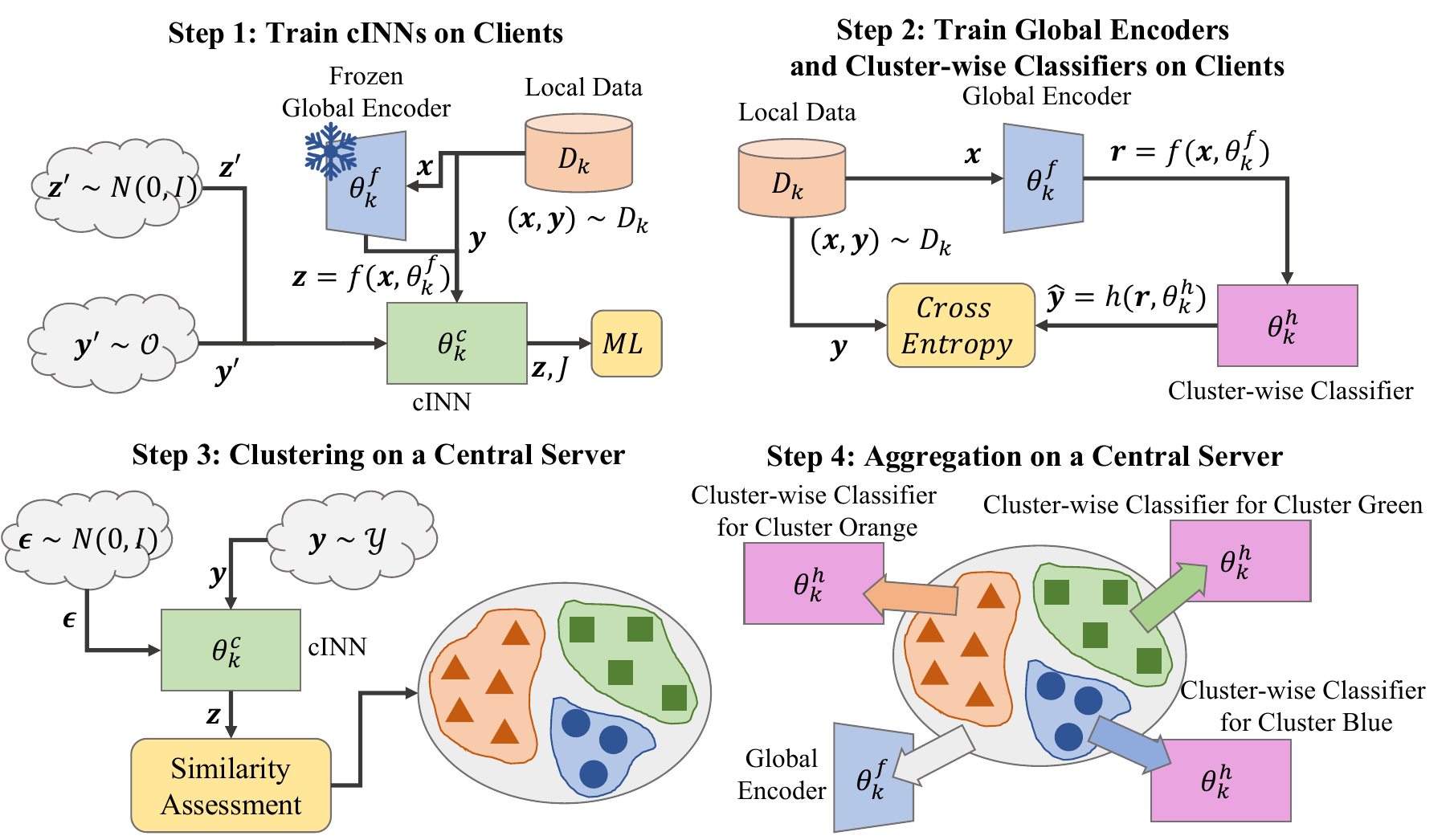}}
\caption{An overview of FCCA.}
\label{fig:fcca-overview}
\end{figure}
Consider a federated learning setting with $N$ clients, where each client $k$ has access to a local dataset $D_k = (D_k^{\mathbf{x}}, D_k^{\mathbf{y}})$, and $(\mathbf{x}_k, \mathbf{y}_k) \sim D_k$ are drawn from the global instance space $\mathcal{X}$ and the global label space $\mathcal{Y}$ respectively. Each client in the setting employs a neural network composed of a global encoder $f: \mathcal{X} \rightarrow \mathcal{Z}$ parameterized by $\theta_k^f$, a cluster-wise classifier $h: \mathcal{Z} \rightarrow \mathcal{Y}$ parameterized by $\theta_k^h$, and a cINN $c: \mathcal{Z} \rightarrow N(0, I)$ parameterized by $\theta_k^c$, where $\mathcal{Z}$ is the latent feature space. Here, we simplify our notation by defining $\theta_k = (\theta_k^f, \theta_k^h, \theta_k^c)$. Moreover, let $\mathbf{z}^{'} \sim N(0, I)$ and $\mathbf{y}_k^{'} \sim \mathcal{O} \in \mathcal{Y} / D_k^{\mathbf{y}}$ signify the out-of-distribution data sample in client $k$. This definition relies on a global label set $\mathcal{Y}$, which is not sensitive to the clients' data content or privacy~\cite{tan2022fedproto}. Nonetheless, it enables the cINN $c$ to accurately represent the out-of-distribution data sample, hereby improving the clustering solution. Finally, let $\mathcal{C} = \{\mathcal{C}_0, \mathcal{C}_1, ..., \mathcal{C}_{M-1}\}$ represents the optimal clustering solution, where $\mathcal{C}_j$ refers to the $j$-th cluster consisting of one or more clients that shares a cluster model of $\theta_j^g$, \textit{i.e.}, $\theta_i = \theta_j^g$ if $i \in \mathcal{C}_j$. We can then formulate the empirical risk minimization of FCCA through Equation~\ref{eq:fcca-objective}.
\begin{equation}
\begin{aligned}
    \argmin_{\mathcal{C}, \theta_0^g, \theta_1^g, ..., \theta_{M-1}^g} \sum_i^{M-1} \sum_{k \in \mathcal{C}_i} \mathcal{L}_k (\mathbf{x}_k; \mathbf{y}_k, \theta_i^g)
\end{aligned}
\label{eq:fcca-objective}
\end{equation}
Here, $\mathcal{L}_k$ represents the local loss function of client $k$, \textit{i.e.,} Cross Entropy Loss. The overall architecture of FCCA is illustrated in Figure~\ref{fig:fcca-overview}. As shown, FCCA comprises four steps: 1) $f$ is frozen for each client $k$, and $c$ is trained using latent features $r=f(\mathbf{x}_k, \theta_k^f)$ and $\mathbf{z}^{'}$ with conditioning data $\mathbf{y}_k$ and $\mathbf{y}_k^{'}$. 2) client $k$ trains its $f$ and $h$ using cross-entropy loss. 3) Gaussian noises $\mathbf{\epsilon}$ and conditioning data $\mathbf{y}_k$ are employed by the central server to reconstruct the distribution of data from the clients. Then, a similarity assessment algorithm is applied to estimate local data distributions of every client for clustering. 4) the central server iterates a new global encoder by aggregating the global encoder of all clients, and generates multiple cluster-wise classifiers based on the clustering solution in step 3.

\subsection{Global encoder and cluster-wise classifier}
Recent works have shown that encoded features supervised by labels follow class-conditional Gaussian distributions~\cite{goodfellow2020generative}. Consequently, these encoded features facilitate the training of class-conditional generators that learns $\mathbf{z}_k$ and selectively preserves only the necessary data, thus ensuring the user's privacy. In FCCA, a global encoder $f$ is utilized to enable consistently mapping from $\mathcal{X}$ to $\mathcal{Z}$ across different clients. The training of $f$ follows standard split federated learning, in which $\theta^{f, t+1} = \sum_{k}^{N-1} p_k^f \theta_{k}^{f, t}$, where $p_k^f$ is the aggregation weight of client $k$ for $f$ that satisfies $\sum_{k}^{N-1} p_k^f = 1$.

To allow for classification of the encoded features, FCCA employs a cluster-wise classifier $h$ that transforms the encoded features to a probability distribution over the classes. $h$ is firstly trained using the encoded features obtained from $f$, and their respective labels. Subsequently, $h$ is aggregated across different clients belonging to the same cluster by $\theta_i^{g, t+1} = \sum_{k \sim \mathcal{C}_i} p_k^g \theta_{k}^{g, t}$, where $p_k^g$ denotes the aggregation weight of client $k$ for $g$ that satisfies $\sum_{k \sim \mathcal{C}_i} p_k^g = 1$.

\subsection{Learning data distributions from global encoder}
\label{sec:cinn}
In FCCA, the local data distribution of each client is captured by a cINN $c$, which maps $\mathcal{Z}$ to $N(0, I)$ based on $\mathbf{y}_k$ and $\theta_k^c$. However, due to data heterogeneity in FL, not all clients have sufficient data to be learned. To tackle this issue, FCCA adopts a data augmentation technique similar to mixup~\cite{zhang2017mixup} to generate synthetic inputs $\mathbf{z}^{'} \sim N(0, I)$ and labels $\mathbf{y}_k^{'} \sim \mathcal{O}$ that represent limited or non-existent data. Essentially, this technique introduces an \texttt{UNKNOWN} label into $\mathcal{Y}$ that is tied with $N(0, I)$.

The rationale behind assigning $N(0, I)$ to the \texttt{UNKNOWN} label is to help the central server measure the confidence of the clients' local data distributions. Detailedly, according to the law of total expectation, $\mathbb{E} \mathbf{a}^\top \mathbf{b} = \mathbb{E} \mathbf{a} \mathbb{E} \mathbf{b}$. As such, $\mathbb{E} \mathbf{z}_1^{'\top} \mathbb{E} \mathbf{z}_2^{'} = 0$ indicates a low level of confidence in the estimation of the clients' local data distributions. Next, taking into account the synthetic inputs and labels, we obtain the loss function for cINN $\mathcal{L}_{cML}$ in FCCA by using the conditional maximum likelihood loss~\cite{ardizzone2019guided}, as follows:
\begin{equation}
    \mathcal{L}_{cML} = \mathbb{E}_{\{\mathbf{z}_k, \mathbf{z}^{'}\}} \frac{||c(\mathbf{z}_k; \mathbf{y}_k, \theta_k^c)||_2^2 + \alpha ||c(\mathbf{z}^{'}; \mathbf{y}_k^{'}, \theta_k^c)||_2^2}{2} - log|J|,
\label{eq:cml-loss-function}
\end{equation}
where $J = \det (\partial c /  \partial \{\mathbf{z}_k, \mathbf{z}^{'}\})$ is the Jacobian determinant evaluated at $\mathbf{z}_k$ and $\mathbf{z}^{'}$, and $\alpha$ is a hyper-parameter that regulates the intensity of the augmentation. The first term of Equation~\ref{eq:cml-loss-function} functions to penalize modes in the training set that have low probability under the given conditioning data and model parameters, and thereby preventing mode collapse.


\subsection{Similarity assessment and clustering}
\label{sec:clustering-algorithm}

After collecting all clients' $\theta_k^c$, the central server will generate a batch of $\mathbf{\epsilon} \sim N(0, I)$ and $\mathbf{y}_k \sim \mathcal{Y}$ to reconstruct $\mathbf{z}_k$ by inverting $c$. However, simply reconstructing $\mathbf{z}_k$ is insufficient for accurate clustering, owing to the presence of poorly learned $\mathbf{z}_k$ resulting from data heterogeneity. To improve the clustering accuracy, we propose to firstly estimate the basic similarity matrix $B$ by Equation~\ref{eq:basic-similarity-matrix}.
\begin{equation}
    B_{\mathbf{y}_k, i, j} = \frac{c^{-1}(\mathbf{\epsilon}_1 ; \mathbf{y}_k, \theta_i^c)^\top c^{-1}(\mathbf{\epsilon}_2 ; \mathbf{y}_k, \theta_j^c)}{||c^{-1}(\mathbf{\epsilon}_1 ; \mathbf{y}_k, \theta_i^c)|| ||c^{-1}(\mathbf{\epsilon}_2 ; \mathbf{y}_k, \theta_j^c)||}
\label{eq:basic-similarity-matrix}
\end{equation}
Here, $[\mathbf{\epsilon}_1$, $\mathbf{\epsilon}_2]$ is split from $\mathbf{\epsilon}$, $i$ and $j$ represent the $i$-th and the $j$-th client, respectively. Then, we can obtain the confidence matrix $P = \max(|B_{\mathbf{y}_k, i, i}|, |B_{\mathbf{y}_k, j, j}|)$.

Next, by fusing $B$ and $P$, the similarity between different clients across different labels, $S$, can be effectively captured with $S_{\mathbf{y}_k, i, j} = B_{\mathbf{y}_k, i, j} P_{\mathbf{y}_k, i, j}$. Finally, we apply the \texttt{K-Means}~\cite{lloyd1982least} algorithm to cluster the data based on the distance matrix $D$, which is obtained from $S$ after dimension reduction based on the arithmetic mean, \textit{i.e.}, $D_{i,j} = \sum_{\mathbf{y}_k}^{\mathcal{Y}} S_{\mathbf{y}_k, i, j} / |S|$, where $|\cdot|$ denotes the length of $\cdot$. 

\begin{algorithm}[tb]
    \caption{Federated cINN Clustering Algorithm (FCCA)}
    \label{alg:fcca-algorithm}
    \textbf{Input}: $\theta$, $\mathbf{x}_k$, $\mathbf{y}_k$, learning rate $\eta_k$\\
    \textbf{Parameter}: number of global epochs $E$, number of local iterations $K$, $N$\\
    \textbf{Output}: $\mathcal{C}$, $\theta^f$, $\theta_0^g$, $\theta_1^g$, ..., $\theta_{M-1}^g$\\
    \textbf{Clients:}
    \begin{algorithmic}[1] 
        \FOR{each client $k$ from $0$ to $N-1$ \textbf{in parallel}}
            \STATE Initialize $\theta_k = \theta$ and Update $\{\theta_k^f, \theta_k^h\}$ for $K$ rounds.
            \FOR{each local iteration $e$ from $0$ to $K-1$}
                \algehlgy{\STATE $\mathbf{z}_k = f(\mathbf{x}_k, \theta_k^f)$}
                \algehlgy{\STATE $\theta_k^c = \theta_k^c - \eta_k \nabla_{\theta_k^c} \mathcal{L}_{cML}(\{\mathbf{z}_k, \mathbf{z}^{'}\}; \mathbf{y}_k, \mathbf{y}_k^{'}, \theta_k^c)$}
            \ENDFOR
            \RETURN $\theta_k = (\theta_k^f, \theta_k^h, \theta_k^c)$
        \ENDFOR
    \end{algorithmic}
    \textbf{The Central Servers}
    \begin{algorithmic}[1] 
        \FOR{each client $k$ from $0$ to $N-1$}
            \STATE receive $\theta_k$ and generate $[\mathbf{\epsilon_1}, \mathbf{\epsilon_2}]$ and $\mathbf{y}_k$ \\
            \algehlgy{\STATE calculate $c^{-1}(\mathbf{\epsilon}_1 ; \mathbf{y}_k, \theta_i^c)$ and $c^{-1}(\mathbf{\epsilon}_2 ; \mathbf{y}_k, \theta_i^c)$}
        \ENDFOR \\
        \algehlgy{\STATE calculate $\mathcal{C}$ by Section~\ref{sec:clustering-algorithm}}
        \STATE $\theta^f = \sum_k^{N-1} p_k^f \theta_k^f$
        \FOR{each cluster $i$ from $0$ to $M-1$}
            \algehlbg{\STATE $\theta_i^g = \sum_{k \in \mathcal{C}_i} p_k^g \theta_k^h$}
        \ENDFOR
        \RETURN $\theta^f$, $\theta_0^g$, $\theta_1^g$, ..., $\theta_{M-1}^g$
    \end{algorithmic}
\end{algorithm}

\subsection{Algorithm and complexity}
The pseudocode for 3SFC is shown in algorithm~\ref{alg:fcca-algorithm}, where the \colorbox{blue!20}{blue} and \colorbox{red!30}{red} code blocks denote the additional computational and memory overheads, respectively. As the algorithm illustrates, the computational complexity equals $\mathcal{O}(2NEK)$ for clients and $\mathcal{O}(2N + 2N^2 + NMT)$ for the central server, and the memory complexity equals $\mathcal{O}(\mathcal{N + M})$ for the central server, where $T$ is the number of iterations in the \texttt{K-Means} clustering process.

\begin{table*}[htb]
  \begin{center}
    \caption{The global top-1 accuracy and personalized top-1 accuracy comparisons (the higher the better) between \colorbox{orange!20}{FCCA and other clustered FL methods} and between \colorbox{cyan!20}{FCCA variants and personalized FL methods} with $M$ = 5.}
    \label{tab:performance-table}%
  \resizebox{1.0\linewidth}{!}{%
    \begin{tabular}{ccccccccccc}
    \toprule
    \toprule
    \multirow{3}[6]{*}{Methods} & \multicolumn{2}{c}{MNIST} & \multicolumn{2}{c}{FMNIST} & \multicolumn{2}{c}{Synthetic} & \multicolumn{2}{c}{Cifar10} & \multicolumn{2}{c}{Cifar100} \\
\cmidrule{2-11}          & \multicolumn{2}{c}{11-layer MLP-based} & \multicolumn{2}{c}{11-layer MLP-based} & \multicolumn{2}{c}{11-layer MLP-based} & \multicolumn{2}{c}{18-layer Conv-based} & \multicolumn{2}{c}{18-layer Conv-based} \\
\cmidrule{2-11}          & Global & Personalized & Global & Personalized & Global & Personalized & Global & Personalized & Global & Personalized \\
    \midrule
    FedAvg~\cite{mcmahan2017communication} & 0.3418$\pm$0.0000 & 0.3064$\pm$0.3002 & 0.3117$\pm$0.0000 & 0.2809$\pm$0.2799 & 0.2573$\pm$0.0000 & 0.2828$\pm$0.2769 & 0.2483$\pm$0.0000 & 0.2050$\pm$0.1831 & 0.0223$\pm$0.0000 & 0.0330$\pm$0.0326 \\
    \midrule
    \cellcolor{orange!20}FL-HC~\cite{briggs2020federated} & 0.3055$\pm$0.1130 & 0.4355$\pm$0.4311 & 0.1742$\pm$0.0729 & 0.2638$\pm$0.2601 & 0.2272$\pm$0.1354 & 0.2877$\pm$0.2817 & 0.1709$\pm$0.0707 & 0.2251$\pm$0.2001 & 0.0194$\pm$0.0087 & 0.0347$\pm$0.0319 \\
    \cellcolor{orange!20}CFL~\cite{sattler2020clustered}   & 0.3403$\pm$0.1245 & 0.3173$\pm$0.3162 & 0.2238$\pm$0.0938 & 0.2795$\pm$0.2744 & 0.2512$\pm$0.0462 & 0.2896$\pm$0.2869 & 0.2111$\pm$0.0479 & 0.1810$\pm$0.1635 & 0.0247$\pm$0.0083 & 0.0333$\pm$0.0307 \\
    \cellcolor{orange!20}IFCA~\cite{ghosh2020efficient}  & \underline{0.8680$\pm$0.0482} & \underline{0.8227$\pm$0.2007} & \underline{0.5409$\pm$0.2130} & \underline{0.4879$\pm$0.3803} & \underline{0.6468$\pm$0.0255} & \underline{0.6983$\pm$0.4246} & \underline{0.3169$\pm$0.1349} & \underline{0.3085$\pm$0.2722} & \underline{0.0372$\pm$0.0348} & \underline{0.0606$\pm$0.0594} \\
    \cellcolor{orange!20}FeSEM~\cite{long2023multi} & 0.2925$\pm$0.0701 & 0.2794$\pm$0.2740 & 0.2833$\pm$0.0723 & 0.2948$\pm$0.2882 & 0.2461$\pm$0.0685 & 0.4102$\pm$0.3948 & 0.2069$\pm$0.0318 & 0.1701$\pm$0.1618 & 0.0179$\pm$0.0042 & 0.0245$\pm$0.0210 \\
    \midrule
    \cellcolor{orange!20}FCCA  & \textbf{0.8800$\pm$0.0136} & \textbf{0.8668$\pm$0.1221} & \textbf{0.7901$\pm$0.0171} & \textbf{0.7617$\pm$0.2340} & \textbf{0.8678$\pm$0.0045} & \textbf{0.8051$\pm$0.2757} & \textbf{0.4585$\pm$0.0253} & \textbf{0.4670$\pm$0.2554} & \textbf{0.0365$\pm$0.0160} & \textbf{0.0654$\pm$0.0925} \\
    \midrule
    \midrule
    \cellcolor{cyan!20}FedPer~\cite{arivazhagan2019federated} & 0.3055$\pm$0.0000 & 0.4355$\pm$0.4310 & 0.2322$\pm$0.0000 & 0.2471$\pm$0.2460 & 0.3545$\pm$0.0000 & 0.6080$\pm$0.4695 & 0.1965$\pm$0.0000 & 0.4307$\pm$0.2890 & 0.0251$\pm$0.0000 & 0.0507$\pm$0.0491 \\
    \cellcolor{cyan!20}FedProx~\cite{li2020federated} & 0.3735$\pm$0.0000 & 0.3226$\pm$0.3160 & 0.3979$\pm$0.0000 & 0.3067$\pm$0.2990 & 0.3236$\pm$0.0000 & 0.3045$\pm$0.2955 & 0.2214$\pm$0.0000 & 0.1935$\pm$0.1685 & 0.0235$\pm$0.0000 & 0.0368$\pm$0.0348 \\
    \cellcolor{cyan!20}PerFedAvg~\cite{fallah2020personalized} & 0.3015$\pm$0.0000 & 0.3182$\pm$0.3091 & 0.2354$\pm$0.0000 & 0.5722$\pm$0.4048 & 0.2429$\pm$0.0000 & 0.2538$\pm$0.2501 & 0.1906$\pm$0.0000 & 0.1802$\pm$0.1554 & 0.0220$\pm$0.0000 & 0.0389$\pm$0.0339 \\
    \midrule
    \cellcolor{cyan!20}FCCA+Per & 0.8755$\pm$0.0110 & 0.8450$\pm$0.1644 & 0.6398$\pm$0.0261 & 0.7074$\pm$0.2780 & 0.6352$\pm$0.0123 & 0.5786$\pm$0.5330 & \underline{0.4430$\pm$0.0318} & \textbf{0.4899$\pm$0.2755} & \underline{0.0385$\pm$0.2477} & \textbf{0.0596$\pm$0.0559} \\
    \cellcolor{cyan!20}FCCA+Prox & \underline{0.9326$\pm$0.0067} & \underline{0.9236$\pm$0.0954} & \underline{0.8315$\pm$0.0101} & \underline{0.7898$\pm$0.2311} & \underline{0.8757$\pm$0.0070} & \textbf{0.7677$\pm$0.3285} & \textbf{0.4809$\pm$0.0214} & \underline{0.4841$\pm$0.2193} & 0.0375$\pm$0.0204 & \underline{0.0549$\pm$0.0518} \\
    \cellcolor{cyan!20}FCCA+Per+Prox & \textbf{0.9329$\pm$0.0078} & \textbf{0.9292$\pm$0.1001} & \textbf{0.8317$\pm$0.0108} & \textbf{0.8321$\pm$0.1546} & \textbf{0.8757$\pm$0.0054} & \underline{0.6814$\pm$0.4503} & 0.4026$\pm$0.0445 & 0.4562$\pm$0.3077 & \textbf{0.0388$\pm$0.0249} & 0.0515$\pm$0.0488 \\
    \bottomrule
    \bottomrule
    \end{tabular}%
    }
  \end{center}
\end{table*}%

\section{Experiments}
\label{sec:experiment}
\textbf{Experimental settings}: All experiments are conducted on $N=100$ clients belonging to one of $M=5$ clusters. The participation ratio is 1.0, the CUDA version is 12.0, the Python version is 3.7.11 and the PyTorch version is 1.10.0. $E=100$, $K=20$, the batch size is $64$, $\eta = 0.01$, and $\alpha = 1.0$. Following the conventions of the community~\cite{sattler2019robust,zhou2021communication}, 5 datasets\footnote{All datasets are publicly available online.} (MNIST, FMNIST, Cifar10, Cifar100 and Synthetic~\cite{caldas2018leaf}) and 2 models (11-layer MLP-based and 18-layer Conv-based) are employed in our experiments. All datasets are split using the Dirichlet distribution~\cite{wang2020tackling} and modified by randomly exchanging labels~\cite{sattler2020clustered} to simulate the Clustered FL setting. For baselines, FedAvg~\cite{mcmahan2017communication}, FL-HC~\cite{briggs2020federated}, CFL~\cite{sattler2020clustered}, IFCA~\cite{ghosh2020efficient} and FeSEM~\cite{long2023multi} are compared with FCCA. Furthermore, to validate that FCCA can be combined with personalized FL algorithms for even better performance, we evaluate the FCCA in combination with FedPer~\cite{arivazhagan2019federated}, FedProx~\cite{li2020federated}, and PerFedAvg~\cite{fallah2020personalized}. \textit{Note that FCCA is not a personalized FL algorithm, the comparisons made with personalized FL algorithms aim to only validate the claim instead of competing with them.} For compared algorithms that have additional hyper-parameters, the values reported in their respective papers are used.

\begin{figure}
     \centering
     \begin{subfigure}[htbp]{0.15\textwidth}
         \centering
         \includegraphics[width=\textwidth]{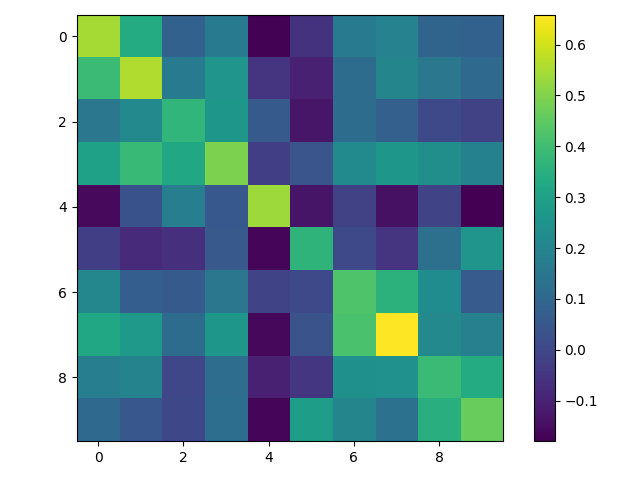}
         \subcaption{$B_{y_k, i, j}$ w/o \texttt{UN-} \texttt{KNOWN}, $E$=10}
     \end{subfigure}
     \begin{subfigure}[htbp]{0.15\textwidth}
         \centering
         \includegraphics[width=\textwidth]{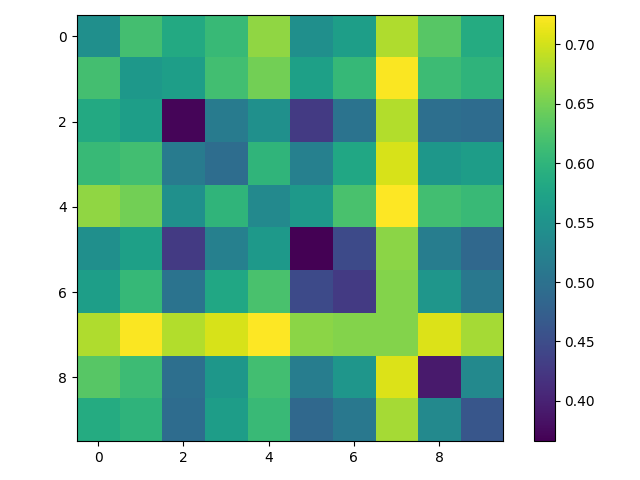}
         \subcaption{$P_{y_k, i, j}$ w/o \texttt{UN-} \texttt{KNOWN}, $E$=10}
     \end{subfigure}
     \begin{subfigure}[htbp]{0.15\textwidth}
         \centering
         \includegraphics[width=\textwidth]{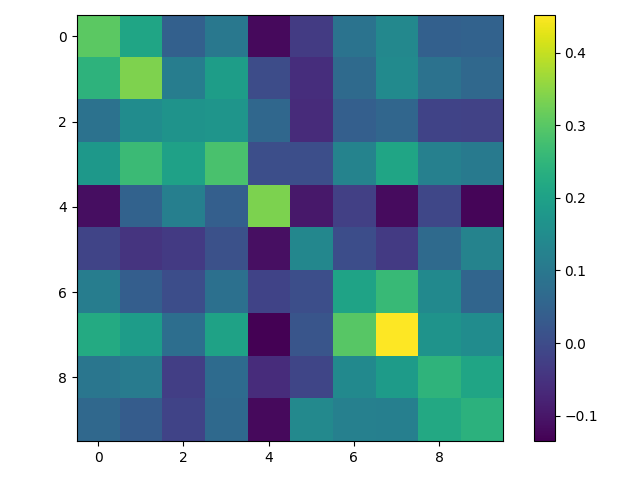}
         \subcaption{$S_{y_k, i, j}$ w/o \texttt{UN-} \texttt{KNOWN}, $E$=10}
         \label{fig:s-wo-unknowns}
    \end{subfigure}
    \\
     \begin{subfigure}[htbp]{0.15\textwidth}
         \centering
         \includegraphics[width=\textwidth]{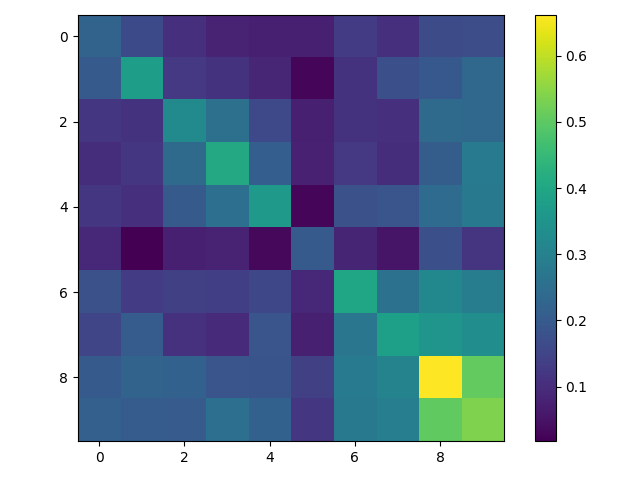}
         \subcaption{$B_{y_k, i, j}$ w/ \texttt{UN-} \texttt{KNOWN}, $E$=10}
         \label{fig:b-w-unknowns}
     \end{subfigure}
     \begin{subfigure}[htbp]{0.15\textwidth}
         \centering
         \includegraphics[width=\textwidth]{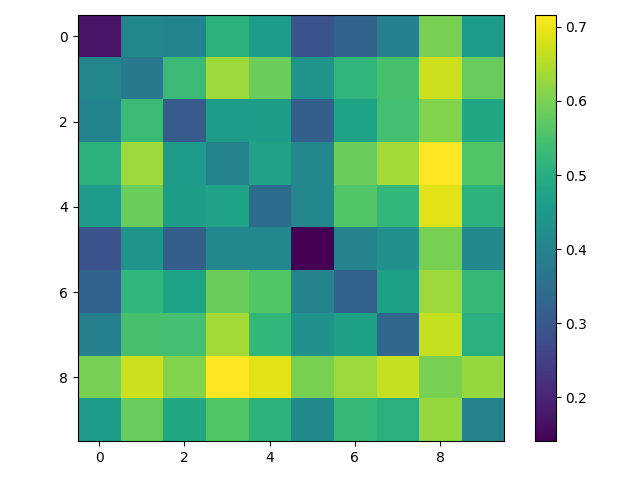}
         \subcaption{$P_{y_k, i, j}$ w/ \texttt{UN-} \texttt{KNOWN}, $E$=10}
     \end{subfigure}
     \begin{subfigure}[htbp]{0.15\textwidth}
         \centering
         \includegraphics[width=\textwidth]{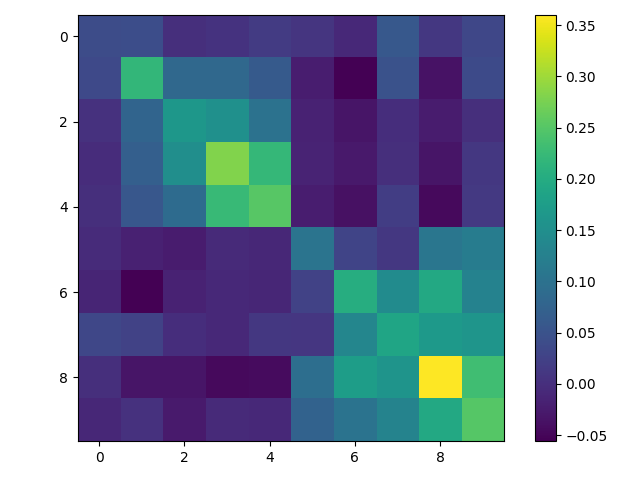}
         \subcaption{$S_{y_k, i, j}$ w/ \texttt{UN-} \texttt{KNOWN}, $E$=10}
         \label{fig:s-w-unknowns}
    \end{subfigure}
    \caption{Similarity assessment of different clients by the central server with $M$ = 2.}
    \label{fig:simil-assess}
\end{figure}

\begin{figure}
     \centering
     \begin{subfigure}[htbp]{0.23\textwidth}
         \centering
         \includegraphics[width=\textwidth]{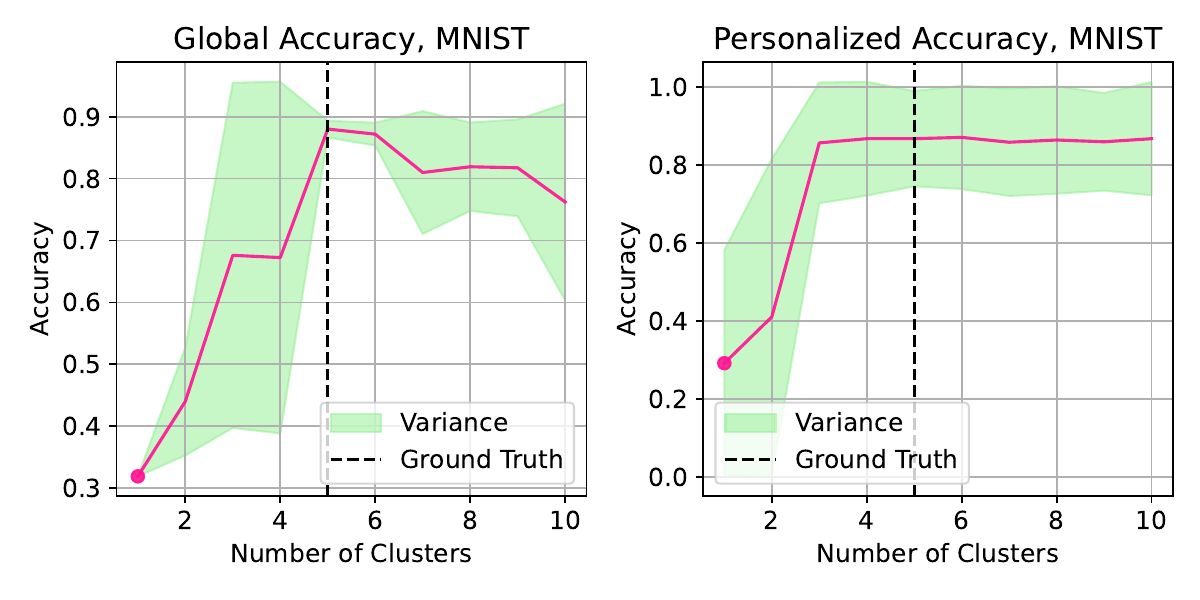}
    \end{subfigure}
    \begin{subfigure}[htbp]{0.23\textwidth}
         \centering
         \includegraphics[width=\textwidth]{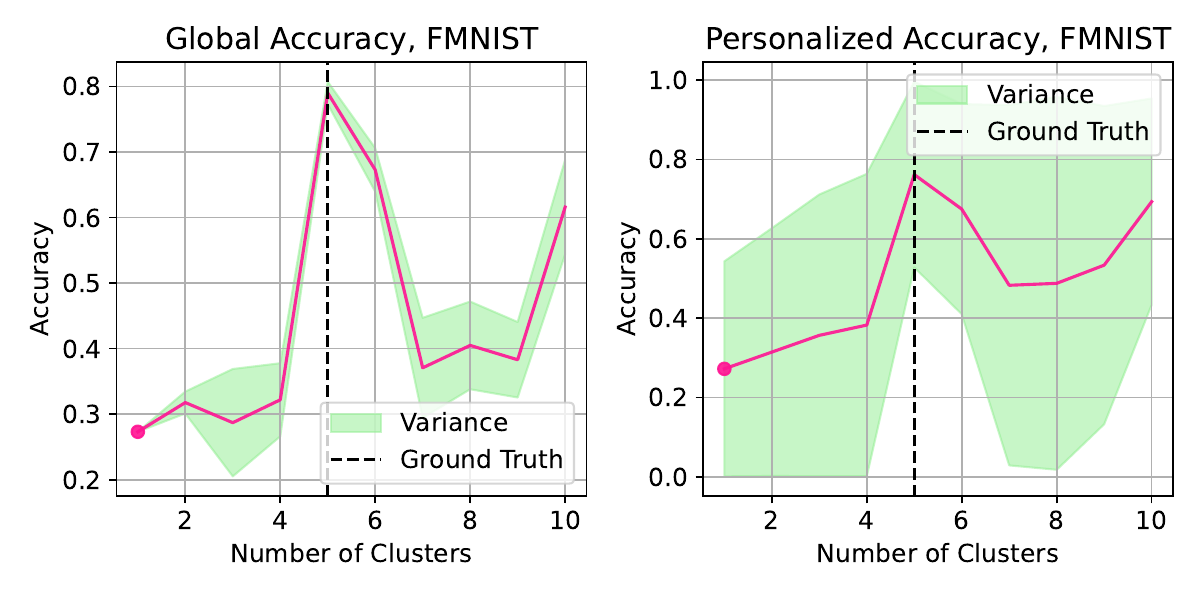}
    \end{subfigure}
    \caption{The final global accuracy and personalized accuracy were evaluated by setting the value of $M$ from 1 to 10, with a ground truth value of 5 for $M$.}
    \label{fig:elbow}
\end{figure}

\textbf{Comparing with existing clustered FL methods}: From the table~\ref{tab:performance-table}, it can be seen that compared to FedAvg, some clustered FL methods achieve lower global and personalized performance due to imprecise and harmful clustering. Conversely, FCCA steadily surpasses all compared clustered FL methods and FedAvg, indicating its robustness and effectiveness in handling non-iid and heterogeneous data.

\textbf{FCCA combined with personalized FL}: FCCA is orthogonal to personalized FL and they can be employed together. Table~\ref{tab:performance-table} shows the performance of standalone personalized FL methods and FCCA combined with personalized FL with $M$ = 5. The results verify the potential of FCCA combined with personalized FL methods for achieving higher global and personalized performance.

\textbf{Validating \texttt{UNKNOWN} labels}: $B$, $P$ and $S$ with and without \texttt{UNKNOWN} labels are compared in Figure~\ref{fig:b-w-unknowns} with $M=2$ (\textit{i.e.}, client 0-4 belong to cluster 0 and client 5-9 belong to cluster 1). It is clear that two clusters become more distinct after applying Section~\ref{sec:clustering-algorithm}. Moreover, Compared to Figure~\ref{fig:s-wo-unknowns}, Figure~\ref{fig:s-w-unknowns} forms clearer cluster, suggesting the noise in $S$ is drastically reduced with \texttt{UNKNOWN} labels.

\textbf{Towards clustering with unknown number of clusters}: As FCCA utilizes \texttt{K-Means} for clustering, it requires the number of clusters to be specified beforehand. From Figure~\ref{fig:elbow}, similar to elbow method~\cite{thorndike1953belongs}, the final global and personalized accuracy of FCCA reaches the highest with minimal variances with the ground truth number of clusters.

\section{Conclusion}
This paper proposes the Federated cINN Clustering Algorithm (FCCA) to overcome the challenge of data heterogeneity in FL. FCCA achieves accurate clustering without cascading errors and mode collapse, while rigorously protecting user privacy. Empirical results show that FCCA is superior to other clustered FL algorithms and can be combined with existing personalized FL algorithms to further boost performance. In future, we will allow FCCA to group clients with unknown $M$ and explore integrating FCCA with other techniques.

\vfill\pagebreak

\end{document}